\documentclass{article}

\usepackage[nonatbib,final]{neurips_2021}

\usepackage{aas_macros}
\usepackage[utf8]{inputenc} 
\usepackage[T1]{fontenc}    
\usepackage{hyperref}       
\usepackage{url}            
\usepackage{booktabs}       
\usepackage{amsfonts}       
\usepackage{nicefrac}       
\usepackage{microtype}      
\usepackage{xcolor}         
\usepackage{subcaption}
\usepackage{graphicx}
\usepackage{comment} 
\usepackage{float}

\definecolor{purple}{rgb}{0.4,0.02,0.58}

\definecolor{darkpastelgreen}{rgb}{0.01, 0.75, 0.24}

\let\subparagraph\paragraph
\usepackage[compact]{titlesec}

\titlespacing{\section}{0pt}{1ex}{1ex}
\titlespacing{\subsection}{0pt}{1ex}{0.5ex}
\titlespacing{\subsubsection}{0pt}{0.5ex}{0.5ex}

\DeclareMathSizes{9}{8.2}{5}{3}
\title{ Physical Benchmarking for AI-Generated \\Cosmic Web}
\author{%
  Xiaofeng Dong \\
  Department of Physics, \\
  University of Chicago, \\
  Chicago, IL 60637, USA \\
  High Energy Physics Division, \\
  Argonne National Laboratory, \\
  Lemont, IL 60439, USA\\
  \texttt{xfdong@uchicago.edu}\\

  \And
  Nesar Ramachandra \\
  Computational Science Division, \\
  Argonne National Laboratory, \\
  Lemont, IL 60439, USA \\
  \texttt{nramachandra@anl.gov}\\
  
  \AND
  Salman Habib \\
  Computational Science Division, \\
  High Energy Physics Division, \\
  Argonne National Laboratory, \\
  Lemont, IL 60439, USA \\
  \texttt{habib@anl.gov}\\
  
  \And
  Katrin Heitmann \\
  High Energy Physics Division, \\
  Argonne National Laboratory, \\
  Lemont, IL 60439, USA \\
  \texttt{heitmann@anl.gov}\\

  \AND
  Michael Buehlmann \\
  High Energy Physics Division, \\
  Argonne National Laboratory, \\
  Lemont, IL 60439, USA \\
  \texttt{mbuehlmann@anl.gov}\\
  
 \And
  Sandeep Madireddy \\
  Mathematics and Computer Science Division,\\
  Argonne National Laboratory,\\
  Lemont, IL 60439, USA \\
  \texttt{smadireddy@anl.gov}\\
}

\begin{document}

\maketitle

\begin{abstract}
The potential of deep learning based image-to-image translations has recently drawn a lot of attention; one intriguing possibility is that of generating cosmological predictions with a drastic reduction in computational cost. Such an effort requires optimization of neural networks with loss functions beyond low-order statistics like pixel-wise mean square error, and validation of results beyond simple visual comparisons and summary statistics. In order to study learning-based cosmological mappings, we choose a tractable analytical prescription -- the Zel'dovich approximation -- modeled using U-Net, a convolutional image translation framework. A comprehensive list of metrics is proposed, including higher-order correlation functions, conservation laws, topological indicators, dynamical robustness, and statistical independence of density fields. We find that the U-Net approach does well with some metrics but has difficulties with others. In addition to validating AI approaches using rigorous physical benchmarks, this study motivates advancements in domain-specific optimization schemes for scientific machine learning.
\end{abstract}

\section{Introduction}

In the current era of `precision cosmology', cosmological N-body simulations play a key role in the study of the large-scale structure of the universe, and are essential to interpreting the unprecedented amount of observational data available from sky surveys \cite{Habib_2016}. In exploring cosmological parameter space, it is computationally challenging to evolve billions and trillions of particles at high enough resolution, even for a single simulation. Since large numbers of simulations are needed for tasks such as covariance estimation \cite{2018arXiv180901669T}, the computing requirements can easily become prohibitive.

Deep learning (DL) has exhibited great potential in learning highly complex functions and mappings, thereby attracting attention in various scientific domains, including cosmology~\cite{Ntampaka2019}. In particular, if DL methods can successfully capture the complexity of cosmic evolution, they can directly address the computational bottleneck, and even eliminate it in some cases. However, issues such as interpretability, as well as the ability of DL models to make sufficiently accurate and physically meaningful predictions, are yet to be fully explored. For example, can DL models, without any domain knowledge of cosmology, capture all the intricacies of nonlinear gravitational clustering? And in what way can AI-based emulators complement standard numerical approaches?

In this ongoing work, we model cosmic evolution using the widely adopted U-Net approach~\cite{He13825}, applying it to a theoretically well-understood prescription, the Zel'dolvich approximation (ZA)~\cite{1970Zeldo}. Since the ZA is a simple dynamical mapping scheme, it has the benefit of providing a clear physical picture while enabling a comprehensive study of the neural network's physical interpretability.

\section{Approximation for structure formation of the Universe}

\paragraph{Dynamical approximation :}
The Zel'dolvich approximation is the first-order solution to the Lagrangian perturbation theory of structure formation
. It describes the motion of mass elements: an initially uniform distribution given by Lagrangian coordinates $\vec{q}$ is displaced by
\begin{equation}
    \vec{x}(t)=\vec{q} + b(t) \vec{S}(\vec{q}),
    \label{eq:co-moving}
\end{equation}
where $\vec{x}(r)$ are the comoving coordinates, $b(t)$ is the linear growth rate of fluctuations and $\vec{S}(\vec{q})= \vec{\nabla} \Phi(\vec{q})$ is the gradient of the initial gravitational potential $\Phi(\vec{q})$. The potential $\Phi(\vec{q})$ is determined by the primordial density fluctuations $\delta(x)$ via the Poisson equation, $\nabla^2 \Phi = 4 \pi G \delta$, where $G$ is the Newtonian constant of gravitation. Initial density perturbations are realizations of a Gaussian random field, which is fully specified by a (given) power spectrum. Uniformly placed particles representing mass elements are then moved according to the ZA. 

\paragraph{Generation of the Dataset:} For an initial power spectrum $P(k)$, the displacement field $\textit{S}(\textit{q})$ is generated using FFT-based techniques, and the particles are transported from a regular grid via ZA displacements. We generate 2000 pairs of ZA-evolved particles at two different time steps of the evolution, with respective scale factors $a_0=0.0464$ (an ``early'' snapshot with redshift $z_0=20.5$) and $a_1=0.215$ (a ``late'' snapshot with redshift $z_1=3.6$). The evolution of the so-called ``cosmic web'' in between these two moments is given by the ZA (from Equation \ref{eq:co-moving}), solely determined by the initial potential gradient. 
The displacement fields at these two times form the basis of the  training/validation datasets. Data is split such that 1000 pairs are used for training, 500 for validation and 500 for testing. Each dataset is generated with a different random seed to ensure their statistical independence.  The dataset consists of $64^3$ N-body particles in a volume of 64 $h^{-1}$Mpc and their displacements in three spatial dimensions.

\subsection{Snapshot-to-snapshot translation using convolutional neural networks}

\begin{figure}
     \centering
         \includegraphics[width=\textwidth]{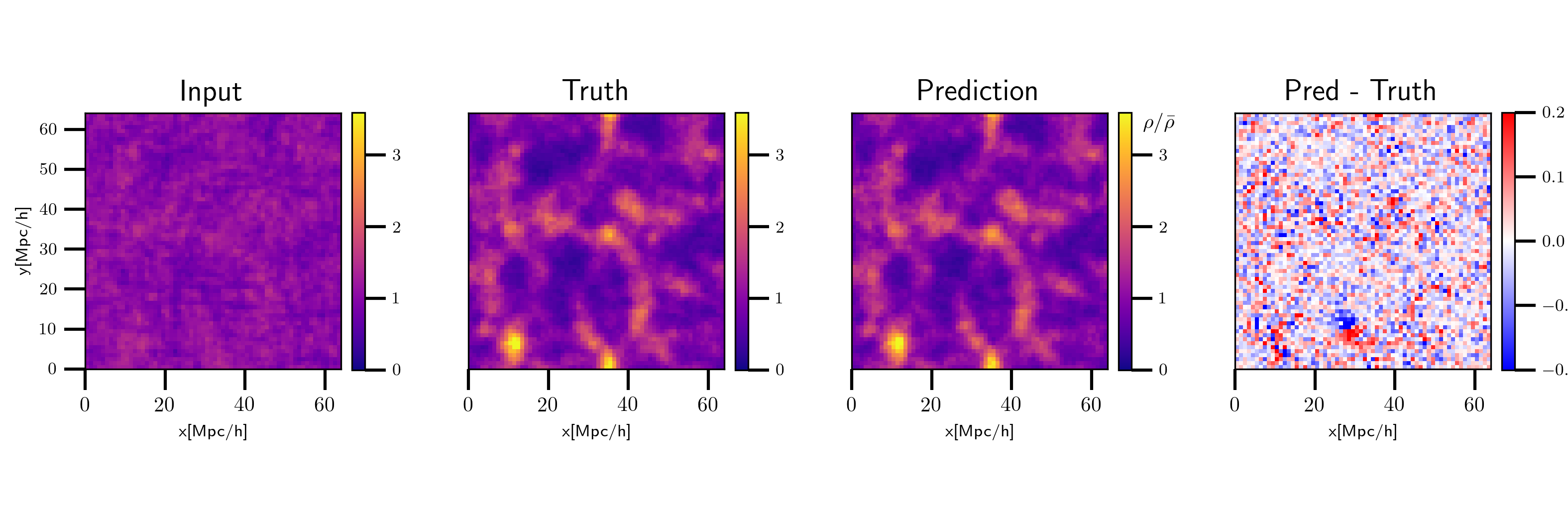}
         \caption{Comparison of projected densities between the predicted cosmic web and ground truth. Densities are derived from the displacement field with a Cloud-In-Cell (CIC) method, and summed over one axis. The `input' panel (first panel) presents the density field projection of the early ($z=z_0$) snapshot. The colorbar in each panel shows the magnitude of the dark matter density $\rho(\textbf{x}$). }
         \label{fig.dens}

\end{figure}

A number of DL-based  unsupervised generative models and supervised interpolation  models have been applied in cosmological data creation~\cite{Ravanbakhsh2016,morningstar2018, Mustafa_2019}. Except for a few applications where the loss functions are tailored to the physical problem, most of the applications are domain-agnostic. That is, all the information about the underlying physics is entirely learned from the training data. These data-driven models have shown good accuracy with validation datasets, albeit with respect to metrics that resemble the loss function.

\paragraph{Architecture:} We choose an image-to-image translation framework~\cite{Ronneberger2015} originally used for biomedical images and recently applied in a cosmological setting~\cite{Giusarma2019, He13825,  Calvo2019, Oliveira2020, Wu2021}. We adopt the 3D U-Net architecture used in \cite{He13825} with 15 convolution and deconvolution layers. The architecture has a contracting path and then an expanding path. A basic block consists of a convolution layer, batch normalization layer and a Rectified Linear Unit (ReLU) layer. Mean squared error between the Lagrangian co-ordinates is used as the loss function, with L2 regularization. We refer the reader to Refs.~\cite{Ronneberger2015,He13825} for further details of the network architecture. 

\paragraph{Model Training:}During the training of the 3D U-Net, a pair of initial-final displacement fields is selected randomly from the dataset as the input, and the neural network is trained to learn the mapping between the $z_0$ and $z_1$ displacement fields. After the training is completed, the model is run on a set of test data, independent from training or validation, in order to compare model predictions with the ground truth (the ZA result). 

\label{gen_inst}

\section{Validation Metrics}
\label{headings}

\begin{figure} 
  \centering
  
 \begin{subfigure}[t]{0.49\textwidth}
 \centering
     \includegraphics[width=\textwidth]{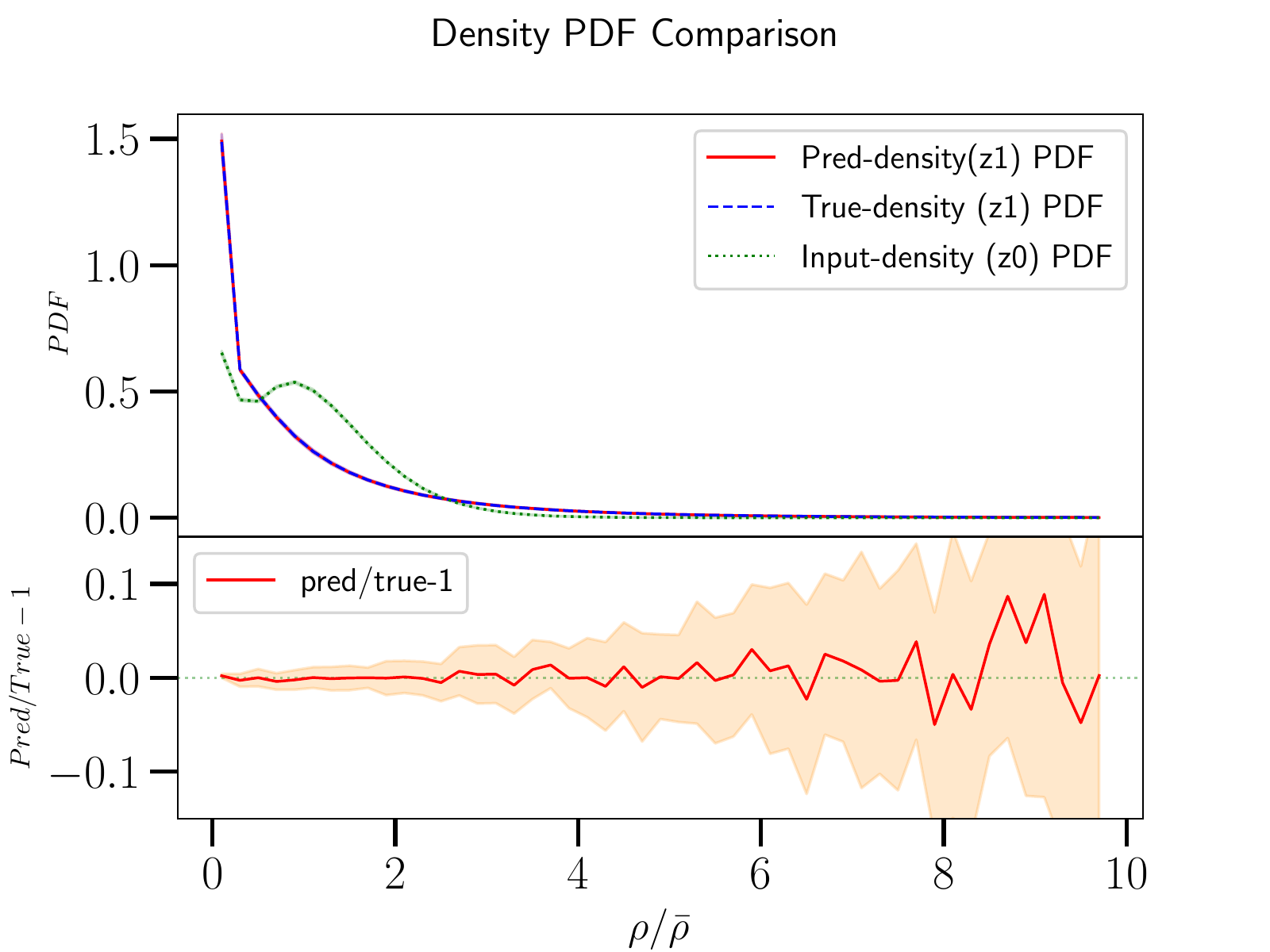}

     \label{fig:densitypdf}
 \end{subfigure}
  \begin{subfigure}[t]{0.49\textwidth}  
 \centering 
      \includegraphics[width=\textwidth]{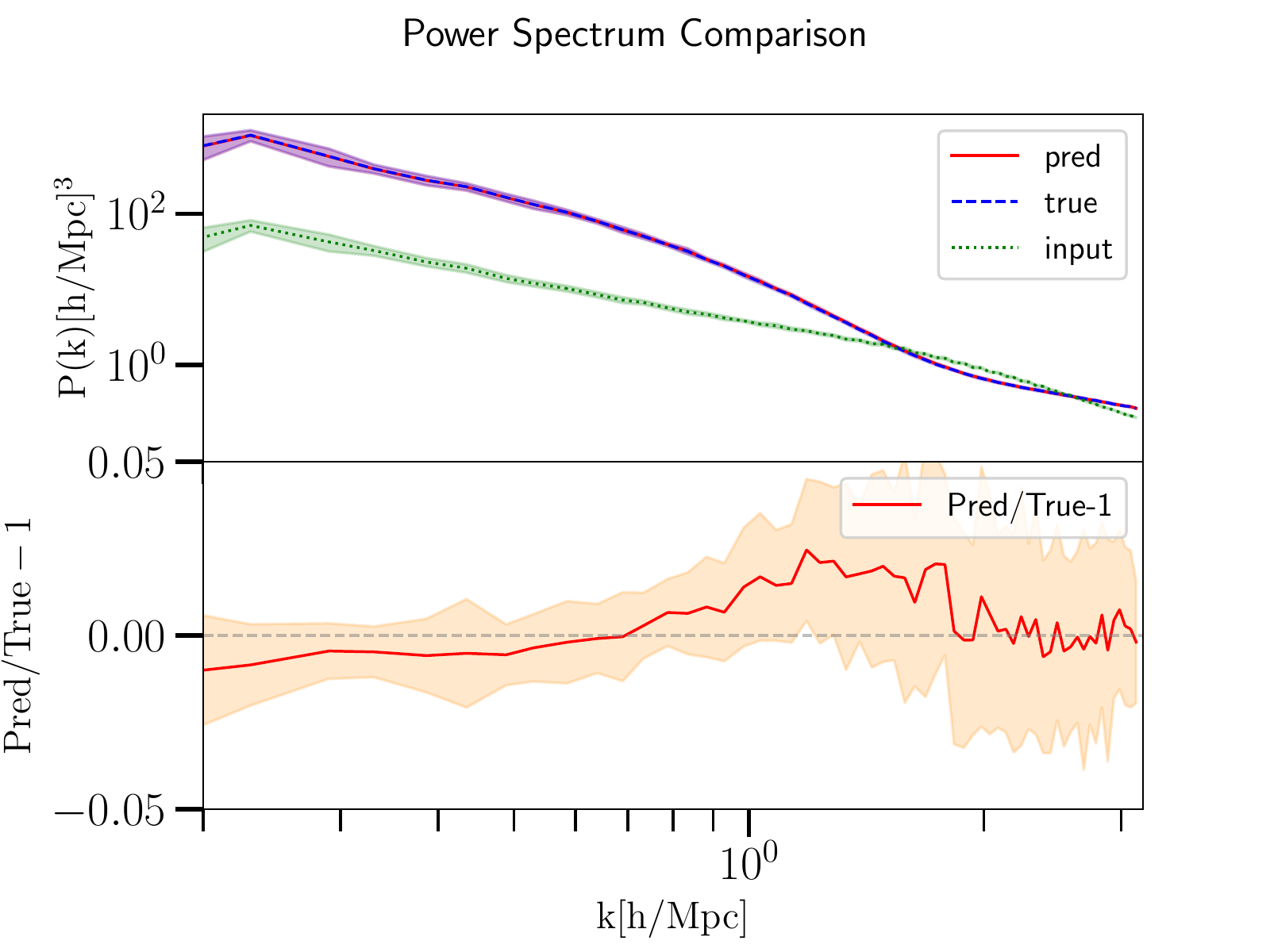}

       \label{fig:pk}
  \end{subfigure}
 
  \begin{subfigure}[t]{0.49\textwidth}
  \centering
     \includegraphics[width=\textwidth]{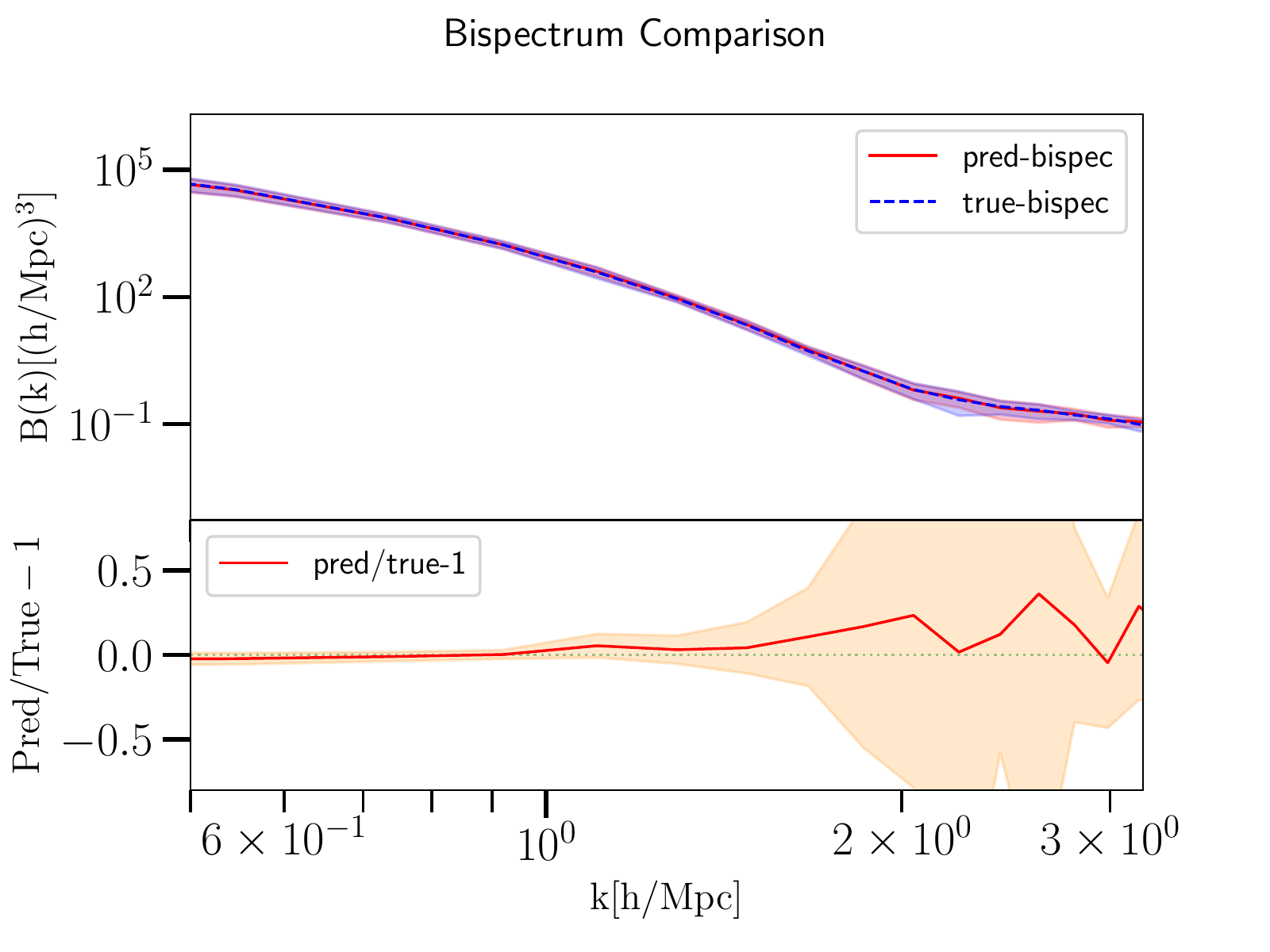}

    \label{fig:bispec}
 \end{subfigure}
  \begin{subfigure}[t]{0.49\textwidth}
  \centering
     \includegraphics[width=\textwidth]{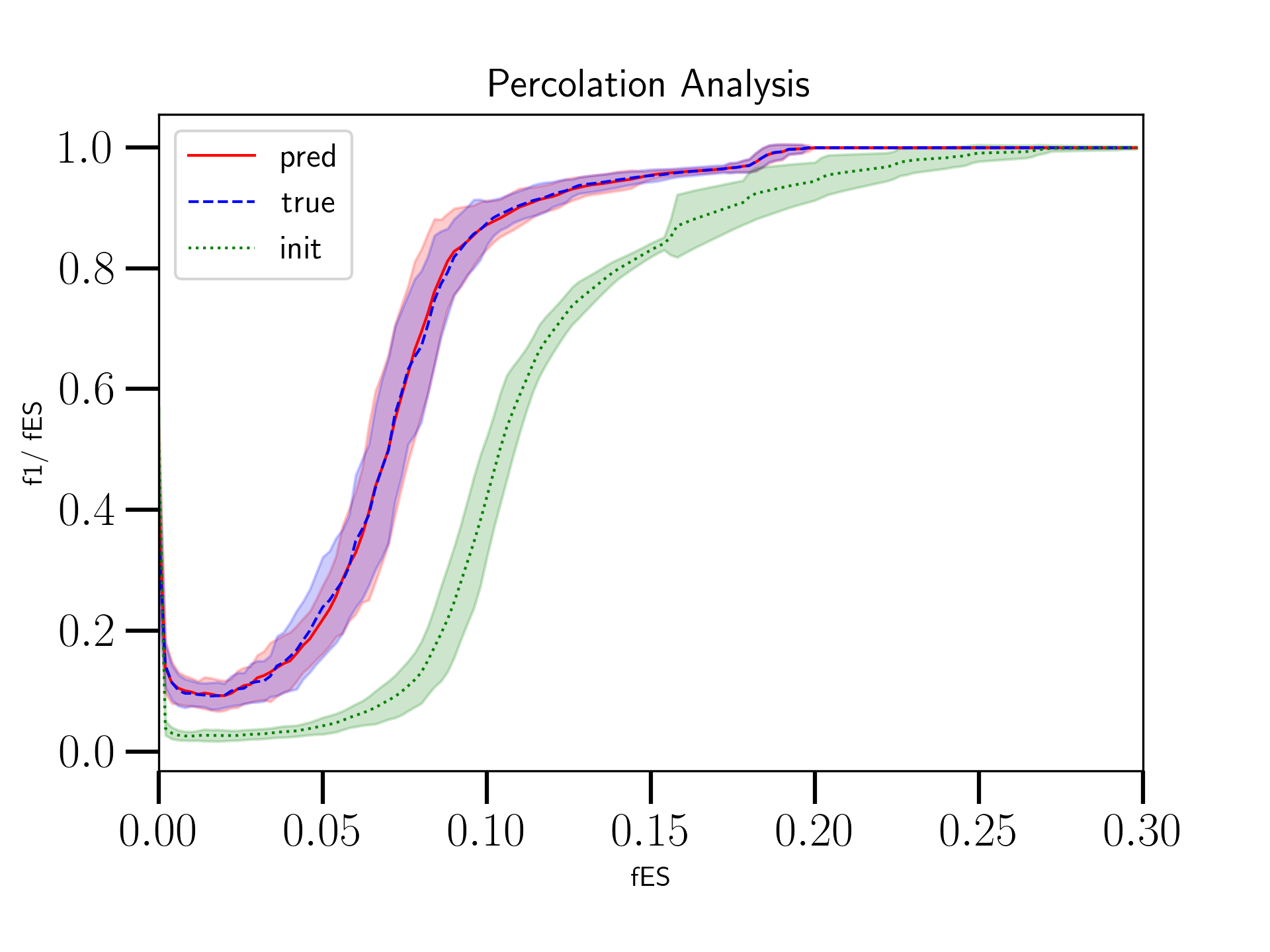}

    \label{fig:percolation}
 \end{subfigure}
\caption{Different physical metrics for assessment of the NN predictions for the ZA-evolved cosmic web. Top left: Density PDF curves for the NN prediction~($z_1$), true~($z_1$) and input~($z_0$) density fields. Top right and bottom left: Matter power spectrum and bispectrum comparison between the predicted density field and ground truth. Bottom right: Percolation transition curve for the truth, prediction and input fields. Shaded area indicates standard deviations over ensemble of 30 realizations.}
\label{fig:physicalmetrics}
\end{figure}

Once the loss has converged (after 10 training epochs), we first visualize the reconstructed cosmic web. Figure~\ref{fig.dens} shows the U-Net generated density (projected along the $x$-axis) of the cosmic density field, demonstrating a reasonably good agreement with the ZA result. We next implement a number of more quantitative tests. 
In particular, we look at measures that quantify the matter distribution in the universe and topological connectivity of large scale structures. Details of these metrics are given below and summarized in Fig.~\ref{fig:physicalmetrics}.


\paragraph{Pixel-wise Comparison:}
The most straightforward comparison is the relative error in predicted densities or displacements. For a density field $\rho(\textbf{x})$, we define a relative error $|\rho(\textbf{x})_{pred}-\rho(\textbf{x})_{truth}|/\rho(\textbf{x})_{truth}$, where $\rho(\textbf{x})_{pred}$ is the U-Net prediction, and $\rho(\textbf{x})_{truth}$ is the ZA result. The last panel of Fig.~\ref{fig.dens} shows the relative difference for one of the realizations.  
In addition, the one-point probability distribution function (PDF) is used to assess the prediction field, which is also an important measure for nonlinear clustering in cosmic structure. The density PDF measurements of prediction and ground truth are shown in the first panel of Fig.~\ref{fig:physicalmetrics}. While relatively unbiased, we note that the density PDF error rises with density, which is concerning if the simulation is expected to capture correct clustering physics.

\paragraph{Matter Power Spectrum:}
The power spectrum is the Fourier transform of the 2-point correlation function in real space, and is widely used as a statistical measure of the density field. The matter overdensity $\delta (\textbf{x})= ({\rho(\textbf{x})-\overline{\rho}})/{\overline{\rho}}$, where $\overline{\rho}$ is the mean density. Writing its Fourier transform dual as $\delta(\textbf{k})$, the power spectrum $P(k)$ is defined by $\langle \delta(\textbf{k}) \delta(\textbf{k}^{\prime}) \rangle= (2\pi)^3 P(k) \delta^3(\textbf{k}-\textbf{k}^{\prime})$.  Measurements of the predicted power spectra and ground truth $P(k)$ are shown in the second panel of Fig.~\ref{fig:physicalmetrics}, where we observe differences of less than 5\%, although there is a $k$-dependent bias in the error behavior. 

\paragraph{Three-Point Correlation -- Bispectrum:}
Apart from two-point correlation functions, higher-order statistics can also be used to characterise the density field, and are especially useful for studying the late stages of structure formation where the evolution is highly nonlinear and non-Gaussian. Adapting similar symbol conventions as above, the bispectrum  $B(k_1,k_2,k_3)$ \cite{hung2019advancing} is defined as:
$\langle \delta(\textbf{k}_1) \delta(\textbf{k}_2)\delta(\textbf{k}_3) \rangle= (2\pi)^3 B(k_1,k_2,k_3) \delta^3(\textbf{k}_1+\textbf{k}_2+\textbf{k}_3)$. Equilateral triangles in $k$ space are selected for our bispectra analysis shown in the third panel of Fig.~\ref{fig:physicalmetrics}. Deviation of the bispectra of the U-Net generated fields with respect to ZA bispectra is considerably higher compared to the lower order metrics of the one-point PDF and matter power spectra.

\paragraph{Topological metrics -- Percolation Analysis:}
Topological statistics can be used to directly address the issue of connectivity of spatial structures, e.g., clusters of galaxies~\cite{Percolation}. For a given density threshold, the volume fraction of the excursion set $f_{ES}$ (i.e., the regions over the threshold) and the volume fraction of the largest region $f_1$ are computed. The behavior of $f_1/f_{ES}$ as a function of $f_{ES}$ (last panel of Fig.~\ref{fig:physicalmetrics}) displays a rapid transition from zero to 1; this signifies a topological shift from isolated clusters to a fully connected field. For the input density field ($z=z_0)$, the percolation transition happens at a different density threshold than $z=z_1$ case, indicating a change in the topological behaviour. For this metric the NN results are very good and in excellent agreement with those from the ZA. 

\begin{figure} 
  \centering
  
  \begin{subfigure}[t]{0.49\textwidth}
  \centering
   \includegraphics[width=\textwidth]{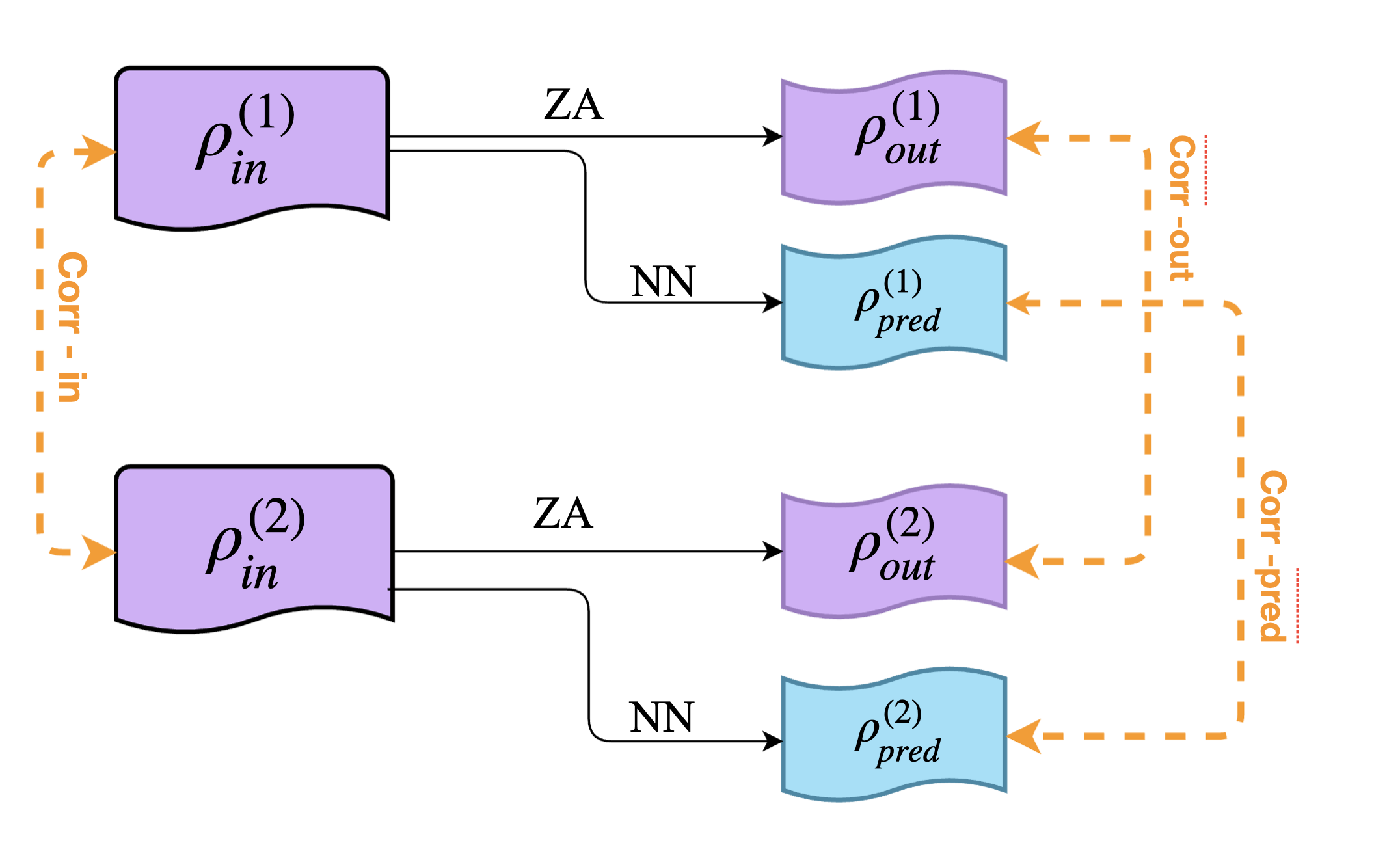}
   \caption{}
   \label{fig:cross-illust}
  \end{subfigure}
  \begin{subfigure}[t]{0.49\textwidth}
  \centering
    \includegraphics[width=\textwidth]{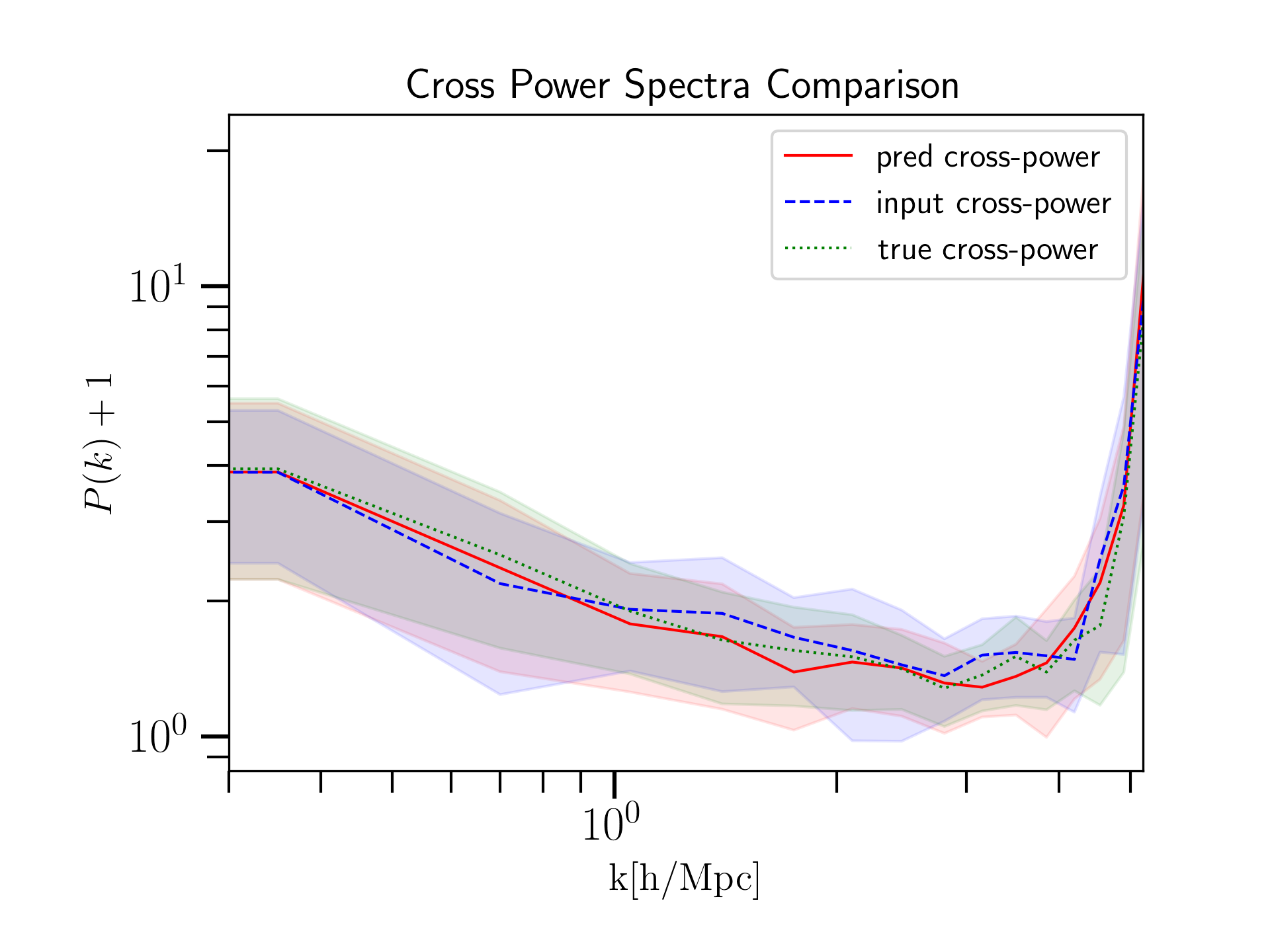}
   \caption{}
  \end{subfigure}
  \caption{Left panel: illustration of the cross-power test, showing the density fields and cross-correlation power spectra computed in between them. Right panel: Mean and standard deviation of the three cross-power spectra (normalized by the auto-power spectra).} \label{fig.crosscorr}
\end{figure}

\paragraph{Cross-power Test:}
The accurate cosmic evolution follows fixed dynamical rules, independent of the initial conditions. However, the neural network emulating the evolution might show discernible bias inherited from the training dataset. Specifically, we need to investigate whether the NN prediction induces otherwise non-existent correlations among the outputs, in contrast to the ZA-evolution of fields, where independent initial conditions should result in independent evolved fields. An example test of a cross-power spectral measurement is shown in  Fig.~\ref{fig.crosscorr}(b), where we generate two initial conditions, independent of the training set, measure their cross-power spectra, then evolve them separately by ZA and NN, and then measure the output cross-power between them again (Fig.~\ref{fig.crosscorr}(a) illustrates this validation route).  From Fig.~\ref{fig.crosscorr}(b), the cross-power shows that, at least for 2-point statistics, there are no hidden correlations among the NN-predicted density fields.

\section{Discussion}

While DL-based translation networks and generative models have demonstrated significant potential, critical tests have to be made before they are applied in high precision scientific fields like cosmology (where we are interested in error metrics aiming at accuracies of better than 1\% \cite{onepercent}). In this work, we have demonstrated a subset of the validation schemes to motivate the scientific machine learning community towards using comprehensive sets of benchmarking tools. Starting from a tractable approximation to the complex dynamics of cosmic structure formation, we demonstrate the use of metrics that were not part of the optimization of the neural network. For instance, topological transitions and bispectra are higher-order emergent features, agreement with which is not trivial in NN-based structure formation models. These checks not only serve as robust validation tools, but also provide a clear direction for designing future cosmic structure emulation schemes. 

We note that although the DL results appear to be promising and close to the true physical results, in order for neural networks to compete with simulations, eventually all errors must be less than 1\% for clustering metrics, and the spatial dynamic range of the 3D NN results must be significantly extended, from a part in a hundred to parts per million. An interesting and long road lies ahead.

\section{Acknowledgements}

The authors would like to thank Ben Gutierrez for useful discussions. Work at Argonne National Laboratory was supported by the U.S.
Department of Energy, Office of High Energy Physics. Argonne,
a U.S. Department of Energy Office of Science Laboratory, is operated by UChicago Argonne LLC under contract no. DE-AC02-
06CH11357. This material is also based upon work supported by the
U.S. Department of Energy, Office of Science, Office of Advanced
Scientific Computing Research and Office of High Energy Physics,
Scientific Discovery through Advanced Computing (SciDAC) program. The training is carried out on Swing, a GPU system at the Laboratory Computing Resource Center (LCRC) of Argonne National Laboratory.
 
{
\small

\bibliography{main}

}

\appendix

\section{Additional metrics}

The metrics studied in this analysis are by no means exhaustive. We list a few other metrics that will be incorporated into our validation scheme in the near future. 

\paragraph{Minkowski Functionals:}
Minkowski Functionals (MFs) are topological descriptors of random fields that are useful in describing their statistical properties \cite{Minkowski}. Given a random field $\rho(\vec{x})$, and its variance $<\rho^2>=\sigma_0^2$, we can consider its excursion sets $\Sigma(\nu)={x>\nu\sigma_0 } $ which consists of all points that exceed the threshold $\nu\sigma_0$. Three MFs measure the area, length of boundary and genus that is characteristic of excursion sets:
$$V_0(\nu) = \frac{1}{A}\int_{\Sigma(\nu)}da $$
$$V_1(\nu) = \frac{1}{4A}\int_{\Sigma(\nu)}dl $$
$$V_2(\nu) = \frac{1}{2\pi A}\int_{\Sigma(\nu)} \mathcal{K} dl $$
where A is the area of the map, $da$ and $dl$ the area and boundary length elements, and $\mathcal{K}$ is the curvature of the boundary.

\paragraph{Dynamical Robustness:}

Apart from the density field, the matter velocity field is of equal significance regarding the correct description of the evolution of the cosmic web, and velocity-density cross correlations are crucial for this measurement. For ZA evolution, explicit reversibility can be also checked to guarantee the dynamical robustness of reconstruction. 


\end{document}